# Optimized General Uniform Quantum State Preparation

## Mark Ariel Levin

University of Maryland, College Park, 20742, Maryland, United States
Email: marklevin.co@gmail.com

**Abstract:** Quantum algorithms for unstructured search problems rely on the preparation of a uniform superposition, traditionally achieved through Hadamard gates. However, this incidentally creates an auxiliary search space consisting of nonsensical answers that do not belong in the search space and reduce the efficiency of the algorithm due to the need to neglect, un-compute, or destructively interfere with them. Previous approaches to removing this auxiliary search space yielded large circuit depth and required the use of ancillary qubits. We have developed an optimized general solver for a circuit that prepares a uniform superposition of any N states while minimizing depth and without the use of ancillary qubits. We show that this algorithm is efficient, especially in its use of two wire gates, and that it has been verified on an IonQ quantum computer and through application to a quantum unstructured search algorithm.

## 1 Introduction

All quantum unstructured search algorithms begin with a uniform superposition representing the search space that is then gradually modified through interference to arrive at a superposition of only correct answers. Typically, uniform state preparation is performed using Hadamard gates, which put a wire into equal superposition of 0 and 1. However, using Hadamard gates to create a uniform superposition means that the size of the search space can only be $2^k$ where k is the number of wires. This means additional nonsensical answers are introduced that decrease search efficiency.

As mentioned in [13] there is need in some cases for a restricted search space. Mukherjee describes a process to find a potential unitary of an initializer for such a restricted search space, namely the Gram-Schmidt procedure. While this works to run this process on a simulator, it cannot be run on an actual quantum computer without first implementing a sequence of gates for this unitary. He gives [2], [9], and [11] as examples to implement the unitary but these do not leverage what is known about the search space and thus do not maximize depth efficiency. In searching for an implementation, we discovered that another potential unitary for the initializer is a Quantum Fourier Transform with an $N^{th}$ root of unity. However, we could not find any implementation of this for any N that is not $2^k$. This process also introduces similar inefficiencies because it imposes unnecessary constraints on the unitary matrix. The only condition necessary for such a state initializer is

$$U|0\rangle_{\lceil \log_2 N \rceil} = \frac{1}{\sqrt{N}} \sum_{n=0}^{N-1} |n_{(2)}\rangle \tag{1}$$

We developed an algorithm that accomplishes this while minimizing 2-wire gate complexity.

The current optimal solver for general non-uniform state-preparation is [6]. For a state represented by d wires, the depth of circuits created by this solver is $O(Nd)$ with constant ancillary wires and $O(\log(Nd))$ with $O(Nd \log N)$ ancillary qubits. However, in scenarios where only a uniform superposition is required, as is often the case in unstructured search problems, our proposed solver offers a notable efficiency advantage. With a reduced circuit depth of $O(\log_2 N)$, fewer 2-wire gates, and the absence of ancillary wires, our solver not only streamlines the preparation process but also minimizes resource requirements, making it a more efficient choice for such applications.

As an example of an unstructured search application of our solver, we will focus hereafter specifically on Graph Coloring Problems. A graph G consists of a set of vertices/nodes V and a set of edges E. Each vertex v has a set of $N_v$ possible colors of which it can exhibit exactly one. An edge between two vertices constitutes a constraint that those two vertices cannot be the same color. A solution to G is a coloring such that each vertex has been assigned a color from among its color-set and all constraints are satisfied.

We have implemented a Grover's Algorithm based approach to such problems as described in [13] to demonstrate an application of our algorithm. For this implementation, each vertex $v \in V$ is represented by $\lceil \log_2 N_v \rceil$ wires. As such, the number of wires needed to solve a given graph is

$$E + 1 + \sum_v^V \lceil \log_2 N_v \rceil \tag{2}$$

Using Hadamard gates for state preparation means a search space that is

$$\prod_v^V 2^{\lceil \log_2 N_v \rceil} \tag{3}$$

which is significantly greater than or equal to the restricted search space

$$\prod_v^V 2^{\log_2 N_v} = \prod_v^V N_v \tag{4}$$

This can mean a significant reduction in efficiency, as the number of repetitions needed is proportional to the square root of the size of the search space. If instead we create a restricted superposition of any given number of states, we can significantly decrease the number of repetitions of Grover's Algorithm that are needed.

For Grover's Algorithm in particular, the sub-circuit for the restricted superposition is needed not only for the initializer but for the mirror of the diffuser in each repetition, making it very important to minimize the number of inefficient 2 wire gates. We believe we have developed an optimal general case solver for the uniform state preparation circuit of any possible N.

## 2 Process

### 2.1 Algorithm

| **Algorithm:** Generate Circuit |
|---|
|     **Input:** Integer N |
|     **Ouput:** Circuit qc |
| **1**   Let j = $\lceil \log_2 N \rceil$ |
| **2**   Initialize an empty circuit qc with j wires |
| **3**   Let i = the number of contiguous 0s at the end of $N_{(2)}$ |
| **4**   Apply a $\frac{\pi}{2}$ **RY** to each of the first i wires of qc |
| **5**   Initialize a list c that will track entangled wires |
| **6**   **for** x = j - 1 **decrementing to** i - 2 exclusive **do** |
| **7**      **if** the x order digit of $(N - 1)_{(2)}$ holds 1 **do** |
| **8**          Apply an **angle**(n,x,c) **RY** rotation to wire x controlled by the last element of c if c has any |
| **9**          Add wire x to the end of c |
| **10**   **for** x = i - 1 **to** j - 1 excusive **do** |
| **11**      **if** x = the last element of c **do** |
| **12**          Remove the last element of c |
| **13**      Apply a $\frac{\pi}{2}$ **CRY** to wire x of qc anti-controlled by the last element of c |
| **14**   **return** qc |

The circuit for the uniform superposition of N states needs j = $\lceil \log_2 N \rceil$ wires. If N is a multiple of $2^i$ for some integer i greater than 0, the first i wires do not need to be entangled because they have an equal desired probability of 1 and 0 independent of the other wires (**line 4**). Once those wires are in superposition, the other wires need to be entangled beginning with the highest order wire. The **angle** function is used to determine the angle of **RY** gate to apply to this wire to achieve the desired probability of 1 and 0. For the subsequent wires, in order of decreasing magnitude, if their corresponding digit in $(N - 1)_{(2)}$ is 1, the **angle** function is used to determine the angle of **CRY** gate to apply to that wire with the most recently rotated wire as the control. This "upward arc" of the algorithm generates the largest element of the superposition.

A subsequent "downward arc" back-fills the remaining elements. First the lowest order wire outside of the first i wires is put into equal superposition anti-controlled by the next lowest order wire that was rotated in the upward arc (recorded in **line 9**). This is because among the states in the superposition where that control wire holds 0 at this point in the process, half should hold 1 in the target wire. This is repeated on each wire in order of increasing magnitude, excluding the highest order wire, each controlled by the next lowest order wire that was entangled in the upward arc.

| **Function:** angle(n,x,c) |
|---|
|     **Input:** Integer n, integer x, and list c |
|     **Ouput:** Float $\theta$ radians |
| **1**   **if** c has elements **do** |
| **2**      $\theta = 2 \sin^{-1} \sqrt{\dfrac{n \bmod 2^x}{n \bmod 2^{c_{last}}}}$ |
| **3**   **else do** |

| 4 | $\theta = 2\sin^{-1}\sqrt{\frac{n\ mod\ 2^x}{n}}$ |
| 5 | **return** $\theta$ |

The desired probability of observing 1 on the highest order wire is the proportion of states in the desired superposition that have 1 on that bit, thus p $= \frac{n\ mod\ 2^x}{n}$ for the highest order wire. For the other wires, the desired probability of observing 1 on the target wire coincident with observing 1 on the control wire is the proportion of states in the superposition that have 1 in the target bit given 1 in the control bit, thus p $= \frac{n\ mod\ 2^x}{n\ mod\ 2^{c_{last}}}$ where $c_{last}$ is the last element of c. The desired magnitude of the 1 state for a given wire is m $= \sqrt{p}$ and to achieve that magnitude we perform an **RY** or **CRY** rotation by $2\sin^{-1}(m)$ depending on whether the wire is of the highest magnitude.

### 2.2 Efficiency

The number of 2-wire gates required to create a uniform superposition for any given N is equal to **count1**(N-1) $- 2$ **maxp**(N) $+ \lceil\log_2 N\rceil - 2$, where **count1**(x)$ is the number of bits holding 1 in the binary representation of x and **maxp**(x) is the maximum integer i such that $x/2^i$ is an integer greater than 1. This can be written mathematically as

$$\sum_{i=1}^{\lceil\log_2 N\rceil-1}\left(\left\lfloor\frac{N\ mod\ 2^i}{2^{i-1}}\right\rfloor - 2\left\lfloor 1-\left(\frac{N}{2^i}\ mod\ 1\right)\right\rfloor\right) + \lceil\log_2 N\rceil - 1 \tag{5}$$

which is $O(\log_2 N)$.

### 2.3 Example 1

Suppose we wanted a superposition of 7 states, see Fig 1. This superposition consists of elements 0 through 6, which are represented in binary as 000, 001, 010, 011, 100, 101, and 110. The wires that represent the value of each bit are labeled in order of decreasing magnitude, $q_2$, $q_1$, $q_0$. Of these 7 states, 3 begin with 1. Thus we want a 3/7 probability of observing 1 on the highest order qubit, $q_2$. The magnitude of the 1 state for this qubit should then be $\sqrt{3/7}$. In order to obtain this magnitude, we perform a **RY** rotation on $q_2$ by $2\sin^{-1}\sqrt{3/7}$. Of the 3 states with 1 in $q_2$, 1 has 1 in $q_1$. Thus we perform a $2\sin^{-1}\sqrt{1/3}$ **CRY** rotation on $q_1$ controlled by $q_2$. Of the 6 states that do not have 1 in both $q_1$ and $q_2$, half have 1 in $q_0$. We perform a $\frac{\pi}{2}$ **CRY** rotation on $q_0$ anti-controlled by $q_1$. Finally, of the 4 states with 0 in $q_2$, half have 1 in $q_1$. The last operation is a $\frac{\pi}{2}$ **CRY** rotation on $q_1$ anti-controlled by $q_2$.

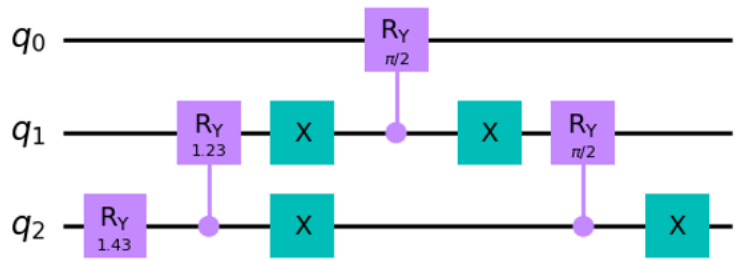

Figure 1: Circuit to initialize a uniform superposition of 7 states.

### 2.4 Example 2

Suppose instead that we wanted a superposition of 22 states, see Fig 2. We follow the same general process with a few modifications. First, because 22 is even, the superposition has an even distribution of 1 and 0 in $q_0$, so it can be put into uniform superposition on its own via a $\frac{\pi}{2}$ **RY** rotation and not entangled. Second, the largest element in the superposition, 21, is represented in binary as 10101. Because $q_3$ in $21_{(2)}$ holds 0, after performing an **RY** on $q_4$, we skip $q_3$ and perform a **CRY** on $q_2$. For the same reason, when we perform the $\frac{\pi}{2}$ anti-controlled CRY rotations on the downward arc of the circuit, $q_3$ is not the control when $q_2$ is rotated. Instead it is controlled by $q_4$.

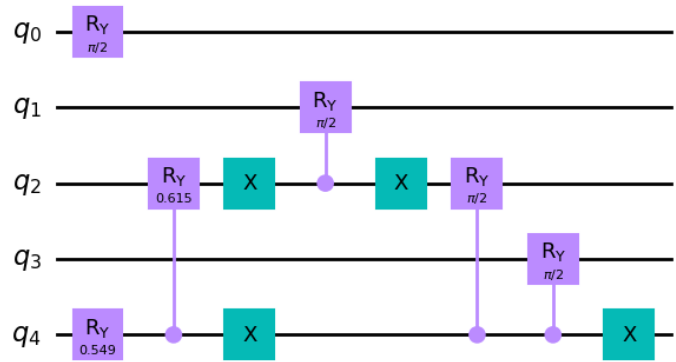

Figure 2: Circuit to initialize a uniform superposition of 22 states.

## 3 Results

### 3.1 Simulator vs Quantum Computer

All results in this section come from the circuit for N=27, see Fig 3.

When run on a noiseless simulator, the algorithm creates a perfect superposition of N states, however, each noisy simulator has a different consistent bias when running, see Fig 4a, likely stemming from inexact rotation angles. Each quantum computer also has a different bias that is consistent across runs in the same calibration, see Fig 4b. However, when a histogram for the frequency of each output is constructed with all of the results from the simulators and the machines, see Fig 4c, the mean bias is very small, though the variance varies greatly between them. This indicates that the bias results from flaws in the hardware, and that the algorithm is sound in principle.

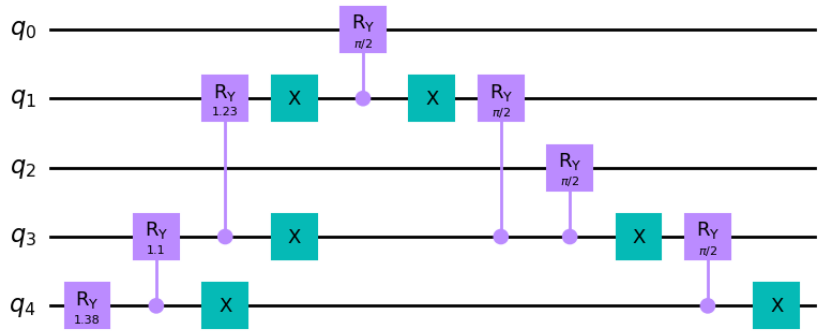

Figure 3: Circuit to initialize a uniform superposition of 27 states.

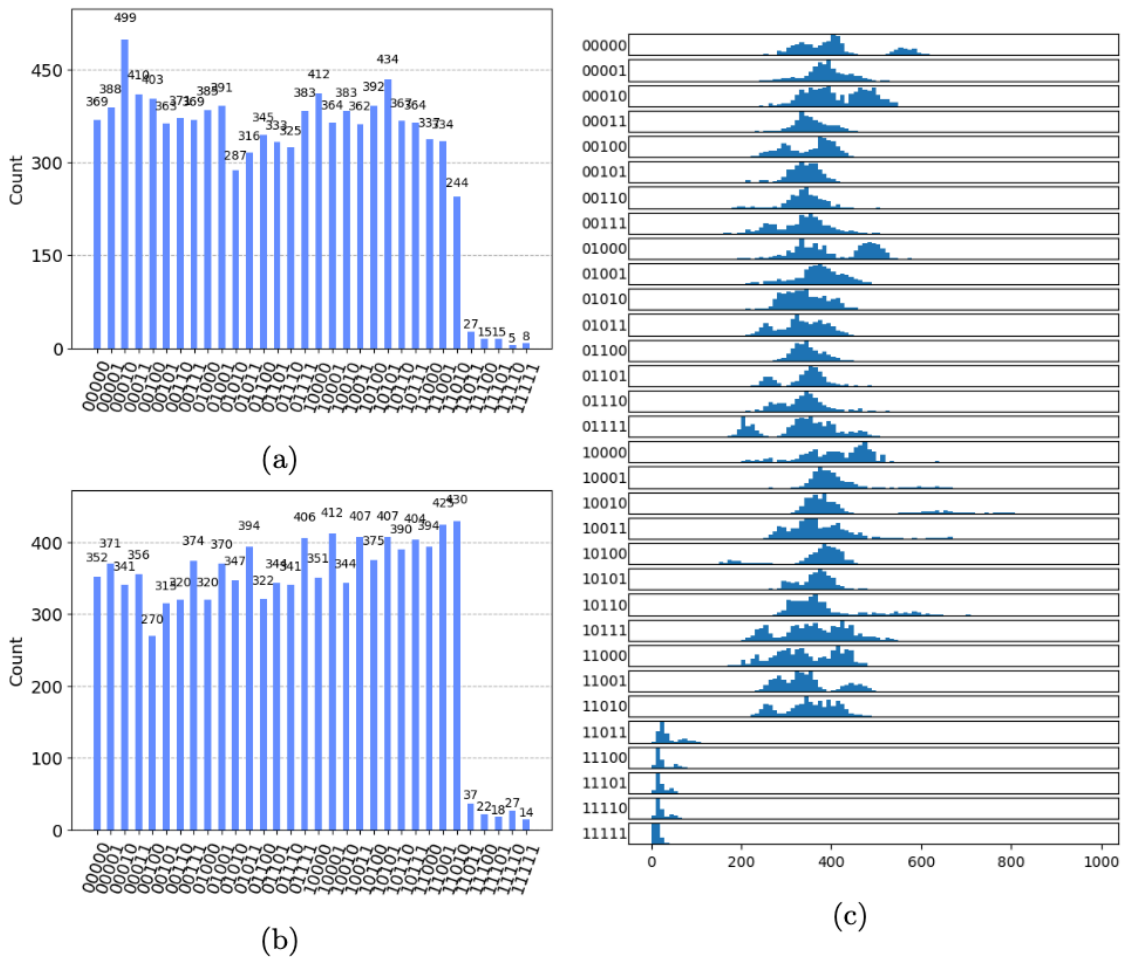

Figure 4: Frequency of each state in the superposition for N=27 with 10,000 shots when run on (a) IonQ's aria noise simulator and (b) their aria machine. (c) Histogram for each state in the superposition for N=27 with 10,000 shots

and results combined from IonQ's harmony and aria noise simulators, their harmony machine, and two different calibrations of their aria machine across 100 runs of each.

It may be of note that the bias when run on a quantum computer is not consistent across calibrations. This was discovered when a group of identical runs on IonQ's harmony machine were spread out over the course of 2 weeks and thus were run under different calibrations, see Fig 5. In this way, the circuits generated by this solver can incidentally be used as a test of the fidelity of a calibration including 2-wire gates.

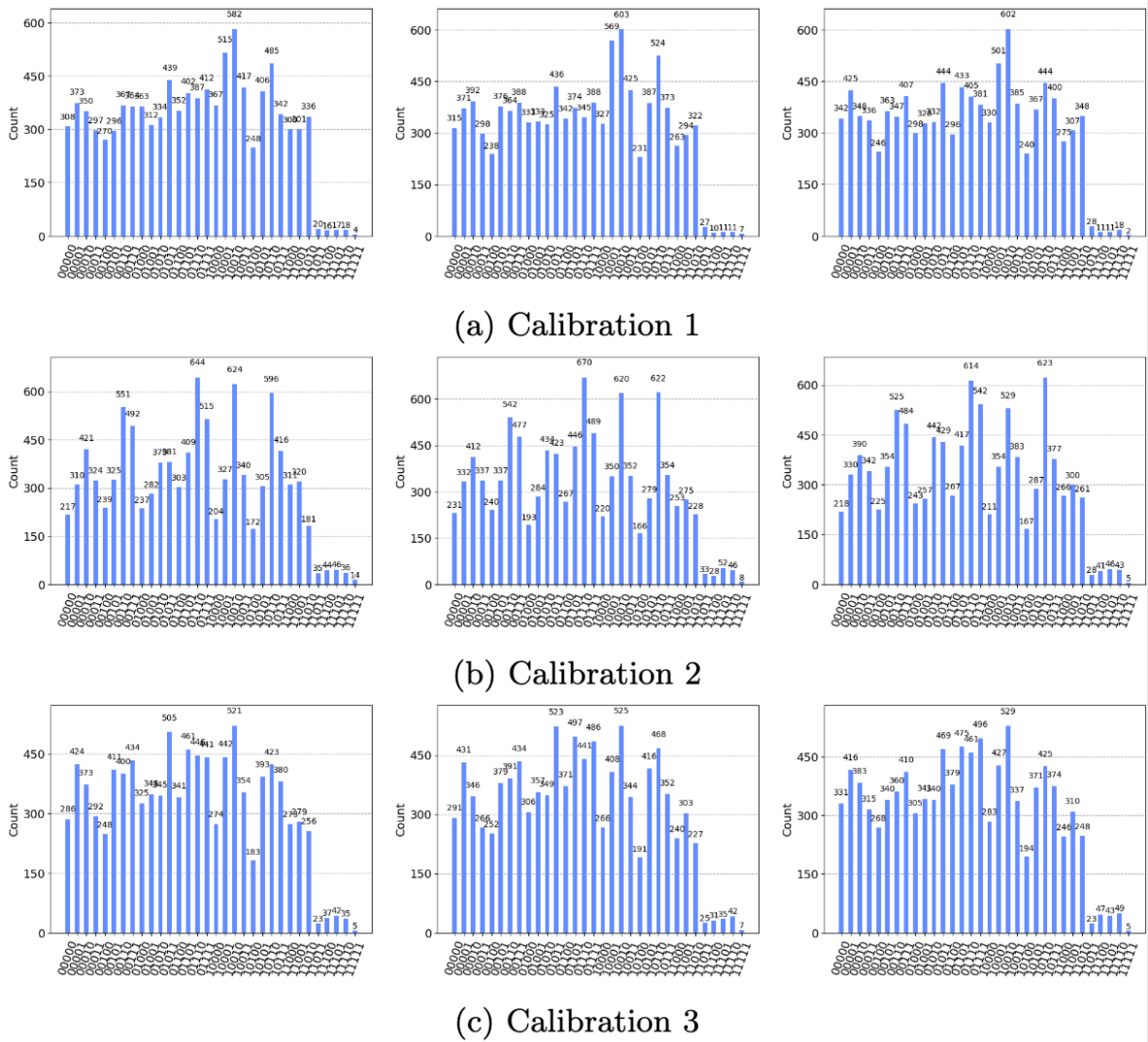

Figure 5: Consistent bias across consecutive runs of N=27 on IonQ Harmony but not across calibrations.

### 3.2 Simulated Modified Grover's Algorithm

When applying this process to Grover's Algorithm for Graph Coloring problems, we saw a remarkable improvement in efficiency that was polynomially proportional to $2^{\lceil \log_2 N \rceil} - N$ and exponentially proportional to the number of vertices/nodes in the graph. This seems to arise naturally from the difference in search space observed in eq (3) and eq (4). On graphs where $2^{\lceil \log_2 N \rceil} - N > 1$, see Fig 6b, Fig 6c , the modified algorithm is more efficient regardless of the number of nodes. However, for certain trivial graph problems where $2^{\lceil \log_2 N \rceil} - N = 1$, a small decrease in efficiency was seen, but this is irrelevant because all such graph problems were trivial and solving is only necessary for non-trivial graph problems. Examples where this is the case include 3-color graphs with 3 or less nodes, see Fig 6a, and 7-color graphs with 7 or less nodes, see Fig 6d. For these graphs, the modified algorithm seems to over-shoot the solution on low-iteration runs and thus requires more runs than the original algorithm to regain a correct solution. However, as the number of nodes increases, the two efficiencies converge until they cross, and the modified algorithm becomes more efficient for the non-trivial graphs. This is clear in Fig 6a, and though Fig 6d ends at 7 nodes, the same trend can be extrapolated. Similar results were observed with both noisy and noise-less simulators.

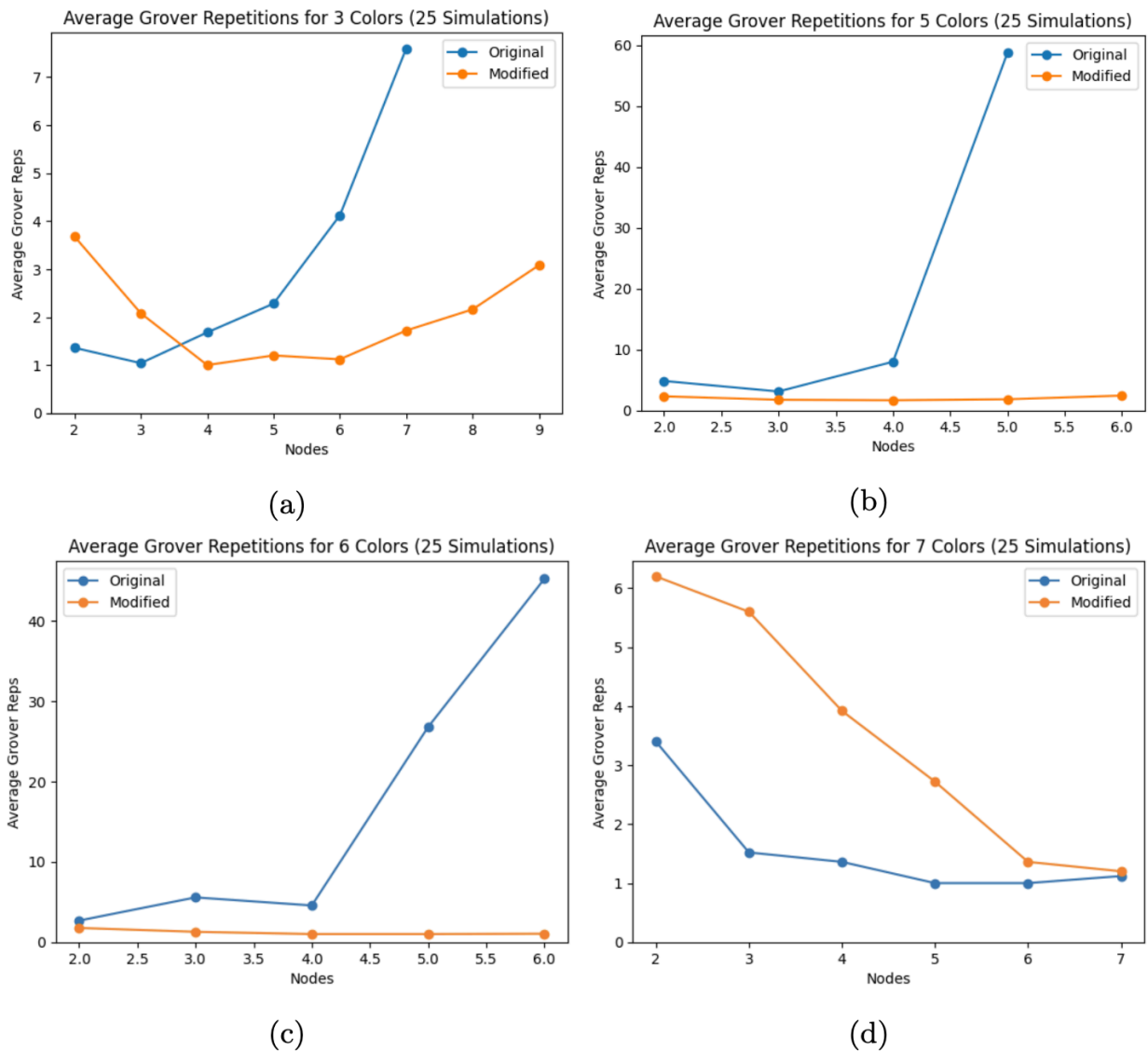

(a)

(b)

(c)

(d)

Figure 6: Average number of Grover Repetitions needed to observe correct answers via a stochastic iterative stepping process as suggested in [8] for a line graph with (a) 3, (b) 5, (c) 6, and (d) 7 colors. These results were simulated on the qiskit Aer simulator with the qiskit FakeVigo noise model.

Unfortunately, trivial graphs predominate in these experiments because of the limitations of the simulator on which they were conducted. When more nodes were attempted than shown in the datapoints, the simulator failed either due to time out, or simply being unable to handle the number of wires required, see eq (2). It is notable that timeout always occurred with fewer or equal nodes for the original version than for the modified version. Graphs where $N = 2^k$ are not included because no modification is made in those cases.

With this modified version of Grover's Algorithm, the initializer and the mirror of the diffuser both have increased depth in each iteration. We have already discussed how we limit the depth of our circuits so that this increased depth is minimized, but the increase in depth is also negligible in comparison to the depth of the oracle and the decomposition of the multi-controlled CNOT in the diffuser, both of which are unchanged by our modification. The data in Fig 6 demonstrate that as a trade-off for this slight increase in the depth of each iteration, we can significantly reduce the number of iterations required to reach a correct solution. Overall, this balances out to significantly decrease the overall depth of Grover's Algorithm.

## 4 Conclusion

We were able to accomplish a significant improvement in the efficiency of quantum unstructured search by eliminating the auxiliary search space created by traditional methods of state preparation while minimizing the number of 2-wire gates required to do so. Although we have only provided the example of a Grover's Algorithm approach to graph coloring problems as a viable application, colleagues of ours have already begun making use of these state preparations to improve the efficiency of quantum walk, machine learning, and other applications. This solver can be used to varying degrees of effect to modify any algorithm that requires a uniform state preparation that is unattainable via simple Hadamard superposition.

**Acknowledgement:**

Thank you to Dr. Franz Klein from University of Maryland for his advisory role, helping me to consider potential applications of my work and for his advice and revisions during the writing of this paper.

**Funding Statement:**

The author received no specific funding for this work.

**Author Contributions:**

Mark Levin was the sole researcher and author of this work.

**Availability of Data and Materials:**

All data in this work can be generated using https://github.com/TheMLevin/GroverGraphSolver.